# Can There Be A Connection Between The Asymmetry Of Angular Distribution Of Electrons In The β *decay* Processes And The Controlled Nuclear Fusion?

## S. I. Fisenko, I. S. Fisenko

"Rusthermosinthes" JSC (The Russian Technologies Venture Fund)
Moskva-City, "Tower 2000"
T. Shevchenko 23A - "B", Moscow, 121151, RUSSIA
Phone: +7 (495) 255-83-64, Fax: +7 (495) 255-83-65
E-mail: StanislavFisenko@yandex.ru

## Abstract

Parity nonconservation in β decay processes is considered to be a fundamental property of weak interactions. Nevertheless, this property can be treated as an anomaly, because in the rest types of the fundamental interactions parity is conserved. Analogously, anomaly in short-duration strong-current pulse discharges is well known. The essence of this phenomenon consists in generation of local high-temperature plasma formations with typical values of its thermodynamic parameters exceeding those related to the central section of a discharge. In this paper, an attempt is undertaken to treat these anomalies as manifestations of the fundamental properties of gravitational emission. Some consequences of this assumption can be tested in β decay experiments as well as in experiments with short-duration *z*-pinch-type pulse discharges.

# 1. Gravitational emission of electrons with a banded spectrum as emission of the same level with electromagnetic emission.

For a mathematical model of interest, which describes a banded spectrum of stationary states of electrons in the proper gravitational field, two aspects are of importance. First. In Einstein's field equations κ is a constant which relates the space-time geometrical properties with the distribution of physical matter, so that the origin of the equations is not connected with the numerical limitation of the κ value. Only the requirement of conformity with the Newtonian Classical Theory of Gravity leads to the small value κ = $8\pi G/c^4$, where $G$, $c$ are, respectively, the Newtonian gravitational constant and the velocity of light. Such requirement follows from the primary concept of the Einstein General Theory of Relativity (**GTR**) as a relativistic generalization of the Newtonian Theory of Gravity. Second. The most general form of relativistic gravitation equations are equations with the Λ term. The limiting transition to weak fields leads to the equation

$$\Delta\Phi = -4\pi\rho G + \Lambda c^2,$$

where Φ is the field scalar potential, ρ is the source density. This circumstance, eventually, is crucial for neglecting the Λ term, because only in this case the GTR is a generalization of the Classical Theory of Gravity. Therefore, the numerical values of κ = $8\pi G/c^4$ and Λ = 0 in the GTR equations are not associated with the origin of the equations, but follow only from the conformity of the GTR with the classical theory.

From the 70's onwards, it became obvious [1] that in the quantum region the numerical value of $G$ is not compatible with the principles of quantum mechanics. In a number of papers [1] (including [2]) it was shown that in the quantum region the coupling constant $K$ ($K \approx 10^{40}\,G$) is acceptable. The essence of the problem of the generalization of relativistic equations on the quantum level was thus outlined: such generalization must match the numerical values of the gravity constants in the quantum and classical regions.

In the development of these results, as a micro-level approximation of Einstein's field equations, a model is proposed, based on the following assumption:

*The gravitational field within the region of localization of an elementary particle having a mass $m_0$ is characterized by the values of the gravity constant K and of the constant Λ that lead to the stationary states of the particle in its proper gravitational field, and the particle stationary states as such are the sources of the gravitational field with the Newtonian gravity constant G.*

The most general approach in the Gravity Theory is the one which takes twisting into account and treats the gravitational field as a gage field, acting on equal terms with other fundamental fields [3]. Such approach lacks in apriority gives no restrictions on the microscopic level. For an elementary spinor source with a mass $m_0$, the set of equations describing its states in the proper gravitational field in accordance with the adopted assumption will have the form

$$\{i\gamma^\mu(\nabla_\mu + \bar{\kappa}\overline{\Psi}\gamma_\mu\gamma_5\Psi\gamma_5) - m_0 c/y\}\Psi = 0 \qquad (1)$$

$$R_{\mu\nu} - \frac{1}{2}g_{\mu\nu}R = -\kappa\{T_{\mu\nu}(E_n) - \mu g_{\mu\nu} + (g_{\mu\nu}S_\alpha S^\alpha - S_\mu S_\nu)\} \qquad (2)$$

$$R(K, \Lambda, E_n, r_n) = R(G, E'_n, r_n) \qquad (3)$$

$$\{i\gamma^\mu\nabla_\mu - m_n c/y\}\Psi' = 0 \qquad (4)$$

$$R_{\mu\nu} - \frac{1}{2}g_{\mu\nu}R = -\kappa'T_{\mu\nu}(E'_n) \qquad (5)$$

The following notations are used throughout the text of the paper: $\kappa = 8\pi K/c^4$, $\kappa' = 8\pi G/c^4$, $E_n$ is the energy of stationary states in the proper gravitational field with the constant $K$, $\Lambda = \kappa\mu$, $r_n$ is the value of the coordinate $r$ satisfying the equilibrium n-state in the proper gravitational field, $\bar{\kappa} = \kappa_0\kappa$, $\kappa_0$ is the dimensionality constant, $S_a = \overline{\Psi}\gamma_a\gamma_5\Psi$, $\nabla_\mu$ is the spinor-coupling covariant derivative independent of twisting, $E'_n$ is the energy state of the particle having a mass $m_n$ (either free of field or in the external field), described by the wave function $\psi'$ in the proper gravitational field with the constant $G$. The rest of the notations are generally adopted in the gravitation theory.

Equations (1) through (5) describe the equilibrium states of particles (stationary states) in the proper gravitational field and define the localization region of the field characterized by the constant $K$ that satisfies the equilibrium state. These stationary states are sources of the field with the constant $G$, and condition (3) provides matching the solution with the gravitational constants $K$ and $G$. The proposed model in the physical aspect is compatible with the principles of quantum mechanics principles, and the gravitational field with the constants $K$ and $\Lambda$ at a certain, quite definite distance specified by the equilibrium state transforms into the filed having the constant $G$ and satisfying, in the weak field limit, the Poisson equation.

The set of equations (1) through (5), first of all, is of interest for the problem of stationary states, i.e., the problem of energy spectrum calculations for an elementary source in the own gravitational field. In this sense it is reasonable to use an analogy with electrodynamics, in particular, with the problem of electron stationary states in the Coulomb field. Transition from the Schrödinger equation to the Klein-Gordon relativistic equations allows taking into account the fine structure of the electron energy spectrum in the Coulomb field, whereas transition to the Dirac equation allows taking into account the relativistic fine structure and the energy level splitting associated with spin-orbital interaction. Using this analogy and the form of equation (1), one can conclude that solution of this equation without the term $\bar{\kappa\Psi}\gamma_\mu\gamma_5\Psi\gamma_5$ gives a spectrum similar to that of the fine structure (similar in the sense of relativism and removal of the principal quantum number degeneracy).. Taking the term $\bar{\kappa\Psi}\gamma_\mu\gamma_5\Psi\gamma_5$ into account, as is noted in [1], is similar to taking into account of the term $\bar{\Psi}o^{\mu\nu}\Psi F_{\mu\nu}$ in the Pauli equation. The latter implies that the solution of the problem of stationary states with twisting taken into account will give a total energy-state spectrum with both the relativistic fine structure and energy state splitting caused by spin-twist interaction taken into account. This fact, being in complete accord with the requirements of the Gauge Theory of Gravity, allows us to believe that the above-stated assumptions concerning the properties of the gravitational field in the quantum region are relevant, in the general case, just to the gravitational field with twists.

Complexity of solving this problem compels us to employ a simpler approximation, namely: energy spectrum calculations in a relativistic fine-structure approximation. In this approximation the problem of the stationary states of an elementary source in the proper gravitational field well be reduced to solving the following equations:

$$f'' + \left(\frac{\nu' - \lambda'}{2} + \frac{2}{r}\right)f' + e^\lambda\left(K_n^2 e^{-\nu} - K_0^2 - \frac{l(l+1)}{r^2}\right)f = 0 \tag{6}$$

$$-e^{-\lambda}\left(\frac{1}{r^2} - \frac{\lambda'}{r}\right) + \frac{1}{r^2} + \Lambda = \beta(2l+1)\left\{f^2\left[e^{-\lambda}K_n^2 + K_0^2 + \frac{l(l+1)}{r^2}\right] + f'^2 e^{-\lambda}\right\} \tag{7}$$

$$-e^{-\lambda}\left(\frac{1}{r^2} + \frac{\nu'}{r}\right) + \frac{1}{r^2} + \Lambda = \beta(2l+1)\left\{f^2\left[K_0^2 - K_n^2 e^{-\nu} + \frac{l(l+1)}{r^2}\right] - e^\lambda f'^2\right\} \tag{8}$$

$$\left\{-\frac{1}{2}(\nu'' + \nu'^2) - (\nu' + \lambda')\left(\frac{\nu'}{4} + \frac{1}{r}\right) + \frac{1}{r^2}(1 + e^\lambda)\right\}_{r=r_n} = 0 \tag{9}$$

$$f(0) = const \ll \infty \tag{10}$$

$$f(r_n) = 0 \tag{11}$$

$$\lambda(0) = \nu(0) = 0 \tag{12}$$

$$\int_0^{r_n} f^2 r^2 dr = 1 \tag{13}$$

Equations (6)—{8} follow from equations (14)—( 15)

$$\left\{-g^{\mu\nu}\frac{\partial}{\partial x_\mu}\frac{\partial}{\partial x_\nu} + g^{\mu\nu}\Gamma^\alpha_{\mu\nu}\frac{\partial}{\partial x_\alpha} - K_0^2\right\}\Psi = 0 \tag{14}$$

$$R_{\mu\nu} - \frac{1}{2} g_{\mu\nu} R = -\kappa \left( T_{\mu\nu} - \mu g_{\mu\nu} \right), \tag{15}$$

after the substitution of $\Psi$ in the form $\Psi = f_{El}(r) Y_{lm}(\theta,\varphi) \exp\left(\dfrac{-iEt}{y}\right)$ into them and specific computations in the central-symmetry field metric with the interval defined by the expression [4]

$$dS^2 = c^2 e^\nu dt^2 - r^2 \left( d\theta^2 + \sin^2\theta d\varphi^2 \right) - e^\lambda dr^2 \tag{16}$$

The following notation is used above: $f_m$ is the radial wave function that describes the states with a definite energy $E$ and the orbital moment $l$ (hereafter the subscripts $El$ are omitted), $Y_{lm}(\theta, \varphi)$ - are spherical functions, $K_n = E_n/\hbar c$, $K_0 = cm_0/\hbar$, $\beta = (\kappa/4\pi)(\hbar/m_0)$.

Condition (9) defines $r_n$, whereas equations (10) through (12) are the boundary conditions and the normalization condition for the function $f$, respectively. Condition (9) in the general case has the form $R(K,r_n) = R(G,r_n)$. Neglecting the proper gravitational field with the constant $G$, we shall write down this condition as $R(K,r_n) = 0$, to which equality (9) actually corresponds.

The right-hand sides of equations (7)—(8) are calculated basing on the general expression for the energy-momentum tensor of the complex scalar field:

$$T_{\mu\nu} = \Psi^+_{,\mu} \Psi_{,\nu} + \Psi^+_{,\nu} \Psi_{,\mu} - \left( \Psi^+_{,\mu} \Psi^{,\mu} - K_0^2 \Psi^+ \Psi \right) \tag{17}$$

The appropriate components $T_{\mu\nu}$ are obtained by summation over the index $m$ with application of characteristic identities for spherical functions [5] after the substitution of $\Psi = f(r) Y_{lm}(\theta,\varphi) \exp\left(\dfrac{-iEt}{y}\right)$ into (17). Even in the simplest approximation the problem of the stationary states of an elementary source in the proper gravitational field is a complicated mathematical problem. It becomes simpler if we confine ourselves to estimating only the energy spectrum. Equation (6) can be reduced in many ways to the equations [6]

$$f' = fP(r) + Q(r)z \qquad\qquad z' = fF(r) + S(r)z \tag{18}$$

This transition implies specific choice of $P, Q, F, S$, such that the conditions

$$P + S + Q'/Q + g = 0 \qquad\qquad FQ + P' + P^2 + Pg + h = 0 \tag{19}$$

should be fulfilled, where $g$ and $h$ correspond to equation. (6) written in the form: $f'' + gf' + hf = Q$. Conditions (19) are satisfied, in particular, by $P, Q, F, S$ written as follows:

$$Q = 1, \qquad P = S = -g/2, \qquad F = \frac{1}{2} g' + \frac{1}{4} g^2 - h \tag{20}$$

Solutions of set (18) will be the functions [6]:

$$f = C\rho(r) \sin\theta(r) \qquad\qquad z = C\rho(r) \cos\theta(r) \tag{21}$$

where $C$ is an arbitrary constant, $\theta(r)$ is the solution of the equation:

$$\theta' = Q\cos^2\theta + (P - S)\sin\theta\cos\theta - F\sin^2\theta \qquad (22)$$

and ρ(r) is found from the formula

$$\rho(r) = \exp\int_0^r \left[ P\sin^2\theta + (Q + F)\sin\theta\cos\theta + S\cos^2\theta \right] dr \qquad (23)$$

In this case, the form of presentation of the solution in polar coordinates makes it possible to determine zeros of the functions *f(r)* at $r = r_n$, with corresponding values of $\theta = n\pi$ (*n* being an integer). As one of the simplest approximations for $v, \lambda$, we shall choose the dependence:

$$e^\nu = e^{-\lambda} = 1 - \frac{\tilde{r}_n}{r + C_1} + \Lambda(r - C_2)^2 + C_3 r \qquad (24)$$

Where řn= (2Kℏ/c³)Kn, $C_1 = \dfrac{\tilde{r}_n}{\Lambda r_n^2}$ , $C_2 = r_n$, $C_3 = \dfrac{\tilde{r}_n}{r_n(r_n + C_1)}$

If we assume that the observed value of the electron rest mass $m_1$ is its mass in the ground stationary state in the proper gravitational field, then $m_o = 4m_1/3$. From dimensionality considerations it follows that energy in the bound state is defined by the expression $\left(\sqrt{Km_0}\right)^2 / r_1 = 0.17 \times 10^6 \times 1.6 \times 10^{-19}$ J, where $r_1$ is the classical electron radius. This leads to the estimate $K \approx 5.1 \times 10^{31}$ Nm²kg⁻² which is later adopted as the starting one. It is known that by use of the dependence $E_0 = \dfrac{e^2}{r_0} = mc^2$ for the impulse we receive the expression $P_i = \dfrac{4}{3}\dfrac{E_0}{c^2}v_i$, that is differentiated by the multiplier 4/3 from the correct expression for the impulse of the particles, the mass of which is $m = \dfrac{E_0}{c^2}$. It is this fact that points at the correctness of the estimates received for the electron, because the lacking part of the energy is in the bound state. From the condition that μ is the electron energy density it follows: μ = 1.1×10³⁰ J/m³, Λ = κμ=4.4×10²⁹ m⁻². From (22) (with the equation for *f(r)* taken into account) it follows:

$$2\theta' = (1 - \overline{F}) + (1 + \overline{F})\cos 2\theta \approx (1 - \overline{F}) \qquad (25)$$

Where $\overline{F} = \dfrac{1}{2}\overline{g}' + \dfrac{1}{4}\overline{g}^2 - \overline{h}$, $\overline{g} = r_n\left(\dfrac{2}{r} + \dfrac{(\nu' - \lambda')}{2}\right)$, $\overline{h} = r_n^2 e^\lambda\left(K_n^2 e^{-\nu} - K_0^2 - \dfrac{l(l+1)}{r^2}\right)$

The integration of equation (25) and substitution of $\theta = \pi n$, $r = r_n$ give the relation between $K_n$ and $r_n$:

$$-2\pi n = -\frac{7}{4} - \frac{r_n K_n^2}{\Lambda^2} \sum_{i=1}^{3} \left\{ A_i \left[ \frac{(r_n + \alpha_i)^2}{2} - 2\alpha_i(r_n + \alpha_i) + \frac{\alpha_i^3}{(r_n + \alpha_i)} + 2C_1(r_n + \alpha_i) + \right. \right.$$

$$\left. + 2C_1 \frac{\alpha_i^2}{r_n + \alpha_i} + \frac{C_2^2 \alpha_i}{r_n + \alpha_i} \right] + B_i \left[ (r_n + \alpha_i) + \alpha_i^2 \frac{1}{r_n + \alpha_i} + \frac{2C_1 \alpha_i}{r_n + \alpha_i} - \frac{C_2^2}{r_n + \alpha_i} \right] \right\} +$$

$$+ \frac{K_0^2 r_n}{\Lambda^2} \sum_{i=1}^{3} A_i'(r_n + \alpha_i) + \frac{r_n l(l+1)}{\Lambda} \left[ d_1 r_n - \frac{C_1 d_2}{r_n} + \sum_{i=1}^{3} a_i(r_n + \alpha_i) \right] - \quad (26)$$

$$- \frac{K_n^2 r_n}{\Lambda^2} \left\{ \sum_{i=1}^{3} \left[ 2\alpha_i^2 A_i - 2\alpha_i B_i - 4C_1 A_i \alpha_i + 2C_1 B_i + C_2^2 A_i + \frac{K_0^2 \Lambda A_i'}{K_n^2}(\alpha_i - C_1) - \right. \right.$$

$$\left. - r_n^2 \Lambda l(l+1) a_i (C_1 - \alpha_i) \right] \ln(r_n + \alpha_i) - r_n \Lambda^{-1} l(l+1)(d_2 + C_1 d_1) \ln r_n \right\}$$

The coefficients entering into equation (26) are coefficients at simple fractions in the expansion of polynomials, required for the integration, wherein $\alpha_i \sim K_n$, $d_2 \sim A_i \sim r_n^{-5}$, $B_i \sim r_n^{-4}$, $A_i' \sim r_n^{-2}$, $a_i \sim r_n^{-4}$, $d_1 = r_n^{-4}$. For eliminating $r_n$ from (26), there exists condition (9) (or the condition exp $v(K,r_n) = 1$ equivalent to it for the approximation employed), but its direct use will complicate the already cumbersome expression (26) still further. At the same time, it easy to note that $r_n \sim 10^{-3} r_{nc}$, where $r_{nc}$ is the Compton wavelength of a particle of the mass $m_n$, and, hence, $r_n \sim 10^{-3} K_n^{-1}$. The relation (26) per se is rather approximate, but, nevertheless, its availability, irrespective of the accuracy of the approximation, implies the existence of an energy spectrum as a consequence of the particle self-interaction with its own gravitational field in the range $r \leq r_n$, where mutually compensating action of the field and the particle takes place. With $l = 0$ the approximate solution (26), with the relation between $r_n$ and $K_n$ taken into account, has the form

$$E_n = E_0 \left(1 + \alpha e^{-\beta n}\right)^{-1}, \quad (27)$$

where $\alpha = 1.65$, $\beta = 1.60$.

The relation (27) is concretized, proceeding from the assumption that the observed value of the electron rest mass is the value of its mass in the grounds stationary state in the proper gravitational field, the values $r_1 = 2.82 \times 10^{-15}$ m, $K_l = 0.41 \times 10^{12}$ m$^{-1}$ giving exact zero of the function by the very definition of the numerical values for $K$ and $\Lambda$.

So, the presented numerical estimates for the electron show that within the range of its localization, with $K \sim 10^{31}$ N m$^2$ kg$^{-2}$ and $\Lambda \sim 10^{29}$ m$^{-2}$, there exists the spectrum of stationary states in the proper gravitational field. The numerical value of $K$ is, certainly, universal for any elementary source, whereas the value of $\Lambda$ is defined by the rest mass of the elementary source. The distance at which the gravitational field with the constant K is localized is less than the Compton wavelength, and for the electron, for example, this value is of the order of its classical radius. At distances larger than this one, the gravitational field is characterized by the constant $G$, i.e., correct transition to Classical GTR holds.

From equation (27) there follow in a rough approximation the numerical values of the stationary state energies: $E_1 = 0.511$ MeV, $E_2 = 0.638$ MeV, ... $E_\infty = 0.681$ MeV.

Quantum transitions over stationary states must lead to the gravitational emission characterized by the constant $K$ with transition energies starting from 127 keV to 170 keV Two circumstances are essential here.

*First.* The correspondence between the electromagnetic and gravitational interaction takes place on replacement of the electric charge e by the gravitational "charge" $m\sqrt{K}$, so that the numerical values K place the electromagnetic and gravitational emission effects on the same level (for instance, the electromagnetic and gravitational bremsstrahlung cross-sections will differ only by the factor 0.16 in the region of coincidence of the emission spectra).

*Second.* The natural width of the energy levels in the above-indicated spectrum of the electron stationary states will be from $10^{-9}$ eV to $10^{-7}$ eV. The small value of the energy level widths, compared to the electron energy spread in real conditions, explains why the gravitational emission effects are not observed as a mass phenomenon in epiphenomena, e.g., in the processes of electron beam bremsstrahlung on targets. A direct confirmation of the presence of the electron stationary states in the own gravitational field with the constant K may be the presence of the lower boundary of nuclear β-decay. Only starting with this boundary β-asymmetry can take place, which is interpreted as parity non-conservation in weak interactions, but is actually only a consequence of the presence of the excited states of electrons in the own gravitational field in β-decay. Beta-asymmetry was observed experimentally only in β-decay of heavy nuclei in magnetic field (for example, $_{27}C^{60}$ in the known experiment carried out by Wu [7]. On light nuclei, such as $_1H^3$, where the β-decay asymmetry already must not take place, similar experiments were not carried out.

## 2. Gravitational Emission in Dense High-Temperature Plasma
### 2.1. Excitation of Gravitational Emission in Plasma

For the above-indicated energies of transitions over stationary states in the own field and the energy level widths, the sole object in which gravitational emission can be realized as a mass phenomenon will be, as follows from the estimates given below, a dense high-temperature plasma.

Using the Born approximation for the bremsstrahlung cross-section, we can write down the expression for the electromagnetic bremsstrahlung per unit of volume per unit of time as

$$Q_e = \frac{32}{3} \frac{z^2 r_0^2}{137} mc^2 \, n_e n_i \frac{\sqrt{2k} \, T_e}{\pi m} = 0.17 \times 10^{-39} z^2 n_e n_i \sqrt{T_e}, \tag{28}$$

where $T_e$, k, $n_i$, $n_e$, m, z, $r_o$ are the electron temperature, Boltzmann's constant, the concentration of the ionic and electronic components, the electron mass, the serial number of the ionic component, the classical electron radius, respectively.

Replacing $r_o$ by $r_g = 2K\,m/c^2$ (which corresponds to replacing the electric charge e by the gravitational charge $m\sqrt{K}$ ), we can use for the gravitational bremsstrahlung the relation

$$Q_g = 0.16 Q_e. \qquad (29)$$

From (28) it follows that in a dense high-temperature plasma with parameters $n_e = n_i = 10^{23}$ m$^{-3}$, $T_e = 10^7$ K, the specific power of the electromagnetic bremsstrahlung is equal to $\approx 0.53 \cdot 10^{10}$ J/m$^3$ s, and the specific power of the gravitational bremsstrahlung is $0.86 \cdot 10^9$ J/m$^3$ s. These values of the plasma parameters, apparently, can be adopted as guide threshold values of an appreciable gravitational emission level, because the relative proportion of the electrons whose energy on the order of the energy of transitions in the own gravitational field, diminishes in accordance with the Maxwellian distribution exponent as $T_e$ decreases.

### 2.2. Amplification of Gravitational Emission in Plasma

For the numerical values of the plasma parameters $T_e = T_i = (!0^7$—$10^8$)K, $n_e = n_i = (10^{23}$—$10^{25}$) m$^{-3}$ the electromagnetic bremsstrahlung spectrum will not change essentially with Compton scattering of electron emission, and the bremsstrahlung itself is a source of emission losses of a high-temperature plasma. The frequencies of this continuous spectrum are on the order of $(10^{18}$—$10^{20})$ s$^{-1}$, while the plasma frequency for the above-cited plasma parameters is $(10^{13}$—$10^{14})$ s$^{-1}$, or 0.1 eV of the energy of emitted quanta.

*The fundamental distinction of the gravitational bremsstrahlung from the electromagnetic bremsstrahlung is the banded spectrum of the gravitational emission, corresponding to the spectrum of the electron stationary states in the own gravitational field.*

The presence of cascade transitions from the upper excited levels to the lower ones will lead to that the electrons, becoming excited in the energy region above 100 keV, will be emitted, mainly, in the eV region, i.e., energy transfer along the spectrum to the low-frequency region will take place. Such energy transfer mechanism can take place only in quenching spontaneous emission from the lower electron energy levels in the own gravitational field, which rules out emission with quantum energy in the keV region. A detailed description of the mechanism of energy transfer along the spectrum will hereafter give its precise numerical characteristics. Nevertheless, undoubtedly, the very fact of its existence, conditioned by the banded character of the spectrum of the gravitational bremsstrahlung, can be asserted. The low-frequency character of the gravitational bremsstrahlung spectrum will lead to its amplification in plasma by virtue of the locking condition $\omega_g \leq 0.5\sqrt{10^3 n_e}$ being fulfilled.

From the standpoint of practical realization of the states of a high-temperature plasma compressed by the emitted gravitational field, two circumstances are of importance.

*First.* Plasma must comprise two components, with multiply charged ions added to hydrogen, these ions being necessary for quenching spontaneous emission of electrons from the ground energy levels in the own gravitational field. For this purpose it is necessary to have ions with the energy levels of electrons close to the energy levels of free excited electrons. Quenching of the lower excited states of the electrons will be particularly effective in the presence of a resonance between the energy of excited electron and the energy of electron excitation in the ion (in the limit, most favorable case — ionization energy). An increase of z increases also the specific power of the gravitational bremsstrahlung, so that on the condition $\omega_g \leq 0.5\sqrt{10^3 n_e}$ being fulfilled, the equality of the gas-kinetic pressure and the radiation pressure

$$k(n_e T_e + n_i T_i) = 0.16(0.17 \cdot 10^{-39} z^2 n_e n_i \sqrt{T_e})\Delta t \tag{30}$$

will take place at $\Delta t = (10^{-6} - 10^{-7})$ s for the permissible parameter values of compressed plasma $n_e = (1 + a) n_i = (10^{25} - 10^{26})$ m$^{-3}$, $a > 2$, $T_e \approx T_e = 10^8$ K, $z > 10$.

*Second.* The necessity of plasma ejection from the region of the magnetic field with the tentative parameters $n_e = (10^{23} - 10^{24})$ m$^{-3}$, $T_e = (10^7 - 10^8)$ K with subsequent energy pumping from the magnetic field region.

**2.3. States of hydrostatic equilibrium of the plasma in the emitted gravitational field**

An analysis of the processes [8, 9] which take place in the known devices for generating stable high-temperature states of a plasma (as well as the absence of encouraging results) suggests that the magnetic field can be used only partially, in the first step for the retention and heating plasma in the process of forming its high-energy state. Further presence of the magnetic field no longer confines the plasma within a limited volume, but destroys this plasma owing to the specific character of motion of charged particles in the magnetic field. A principal solution of the problem is a method of confining of an already heated plasma in an emitted gravitational field in a second step, after the plasma has been compressed, heated and retained during this period by the magnetic field. As follows from the above-stated, under any circumstances plasma must be injected from the magnetic field region, but with subsequent pumping of energy from the region of the plasma found in the magnetic field.

The fulfillment of the above-cited conditions (in principle, irrespective of a particular scheme of the apparatus in which these conditions are realized), and namely:

a) generation of a dense high-temperature plasma from hydrogen and isotopes thereof with the aid of pulsed heavy-current discharges;

b) injection of the plasma from the area of a magnetic field with parameters corresponding to the conditions of gravitational emission of electrons with a banded energy spectrum;

c) energy transfer along the spectrum performed by cascade transition into the long wavelength region of eV-energy to the state of locking and amplification of the gravitational emission and simultaneous compression to the states of hydrostatic equilibrium, and in the formation of the states, mentioned for the stage c) in the composition of a working gas multielectron atoms are used for quenching the spontaneous gravitational emission from the ground energy levels of the keV-region electron in its own gravitational field, solves solely the problem of attaining hydrostatic equilibrium states of plasma. The use of a multielectron gas (carbon) as the additive to hydrogen leads to the realization of nuclear fusion reaction conditions, since carbon will simultaneously will act as a catalyst required for the nuclear fusion reaction [10].

Another variant of nuclear fusion in compositions with multielectron atoms, such as xenon, krypton (and allied elements) is the use as the light component of deuterium or a deuterium-tritium mixture (the last-mentioned variant is the most cumbersome one because of the forming of neutrons during the fusion).

**Conclusion**

1. The approximation of Einstein's relativistic gravitation equations is examined, wherein the values of gravitational constant K and constant $\Lambda$ in the region of elementary particles localization are defined in such a way that this leads to stationary states of the particles in their own gravitational field, and the particles stationary states themselves are the sources of the field with Newtonian gravitational constant G.

2. The consequences of the quantum character of gravitational interaction, which are available for experimental testing, are:

   a) asymmetry of electrons scattering on the energies in $\beta$-decay in the strong magnetic field;
   b) state of hydrostatic equilibrium of dense high-temperature plasma in the emitted gravitational field.